# Plasmonic Double-hole Bull's Eye Nano-antenna for Far-field Polarization Control

*Abbas Ghaffari, Somayeh Kashani, Jiazhen Li, Paschalis Gkoupidenis, Robert Riehn, Qing Gu*[*]

ABSTRACT: Plasmonic polarization conversion offers significant advantages over conventional methods, including a smaller device footprint and easier integration into photonic circuits. In this work, we numerically and experimentally investigate the polarization conversion properties of a plasmonic double-hole structure surrounded by circular nanograting, i.e., a bull's eye antenna. Using a combination of polarimetric imaging via back focal plane (BFP) microscopy and the Stokes parameter analysis, we demonstrate the functionality of our structure as a miniature on-chip polarization converter. Our results show that this nanostructure enables complex polarization transformations, including converting linear to circular polarization and vice versa. The polarization conversion efficiency is found to be dependent on the periodicity of the circular gratings and is particularly pronounced in the central region of the Fourier space. Moreover, the strong asymmetric scattering leads to distinctive patterns in the Stokes parameters across various incident polarization states. This work provides insights into the plasmonic manipulation of light polarization at the nanoscale, with potential applications in miniature on-chip polarization convertors, polarization-controlled emitters, and advanced sensing technologies.

# Introduction

Over the past two decades, significant advancements have been made in the fields of nanophotonics and plasmonics thanks to the rapid development of novel materials and nanofabrication technologies. These advancements have enabled the precise manipulation of light at the nanoscale, unlocking a wide range of applications, from photocatalysis and electrocatalysis to various optoelectronic devices for sensing, energy harvesting, and bio-treatment. Additionally, the versatile design of metasurface, chiral structures, and electromagnetic field localization have led to a deeper understanding and control over the fundamental properties of light[1,2]. In particular, it has been demonstrated that plasmonic structures can enable polarization conversion, allowing for the precise control of amplitude, phase, and polarization states of electromagnetic waves through engineered geometries, resonances, and near-field interactions. Plasmonic polarization conversion offers significant advantages over conventional methods, including a smaller device footprint, easier integration into photonic circuits, higher wavelength selectivity, and more tunable polarization control.

State-of-the-art plasmonic polarization converters have utilized various approaches to manipulate or control the polarization state of light. These include plasmonic chiral nanostructures[3,4] and metasurfaces with tailored[5,6] chiroptical properties, plasmonic nanostructures integrated with quantum emitters for polarization-dependent emission[7], ultrafast polarization switching in nonlinear plasmonic metasurfaces[8], and polarization control through thermal or spin effects[9,10]. Among these structures, the plasmonic bull's eye antenna - consisting of a central subwavelength aperture surrounded by concentric periodic grating etched in a metal film - is particularly noteworthy because of its simplicity and effectiveness[11–15]. This structure allows for precise control over beam shape and radiation direction by coupling localized and propagating surface plasmons[12,13,16]. The performance of the bull's eye antenna can be tailored by adjusting antenna geometry, such as periodicity, and controlling the optical properties of the antenna environment, such as the refractive index of the surrounding material, making this design valuable for various nanophotonic applications, including efficient photon collection, highly directional room-temperature single photon sources[7,17–22].

As demonstrated by Lezec et al. and further studied by others[12,13,23], the bull's eye antenna and periodic structure can enable efficient coupling of incident light to surface plasmon polaritons (SPPs), enhancing the transmission coefficient through the aperture by orders of magnitude. Furthermore, these structures allow control over the directionality of transmitted light and concentrate electromagnetic energy at the central aperture. This beaming effect results from the coherent interference between directly transmitted light and SPP-scattered light from the grating, where the radial propagation wavelength through the grating matches the grating period, and all scattering events from the grating are in phase[13,24]. Consequently, plasmonic bull's eye structures have been used to enhance and control the unidirectional and poorly polarized emission from single photon sources[25–27]. When used with light emitting materials, refs [25, 26, 27, 28] showed that this structure enables significant fluorescence enhancement with highly directional emission and narrow beam divergence angles[25,28]. When a quantum emitter is placed in the central aperture, its emission couples to SPPs that propagate radially outward along the metal surface, resulting in more photons being directed toward the collection optics, which leads to a large improvement in collection efficiency.

On the other hand, polarization, i.e., the vector characteristic of light, plays a pivotal role in light-matter interaction, which can also be controlled through various plasmonic designs[12,29–31]. Various experimental approaches have been employed to study polarization effects in plasmonic structures, including polarization-resolved measurements like Stokes and Mueller matrix polarimetry[32], K-space polarimetry for far-field angular distribution[11], and near-field mapping techniques[33,34]. Recently, polarimetry imaging has drawn increasing attention for its ability to significantly enhance image contrast, particularly when imaging objects are in a scattering media[35]. Furthermore, it holds substantial potential for applications in biomedical science and environmental research[32,36].

We previously demonstrated that the bull's eye antenna can function as a refractive index sensor by efficiently coupling SPPs to the far field and creating a distinct interference pattern that is highly sensitive to refractive index changes within the subwavelength aperture[26]. In this work, we exploit the use of bull's eye antenna to manipulate the polarization characteristics of the transmitted light, with an aperture consisting of a pair of plasmonic holes. We further explore its function as a polarization converter. We use back focal plane (BFP) imaging to characterize the spatial distribution of the transmitted light transmitted through the structure and measure the Stokes parameters at all points collected by the microscope objective.

**Results and discussion**

The proposed nanoplasmonic polarization converter is based on a pair of nanoholes embedded in the center of a bull's eye antenna with six concentric circular corrugations patterned on a 160 nm-thick gold film. The grating periodicity, ranging from 700 to 800 nm, is used to modify the polarization of the incident beam. The gold film is supported by a fused silica substrate with a chromium adhesion layer, and the structure is fabricated using focused ion beam (FIB) milling.

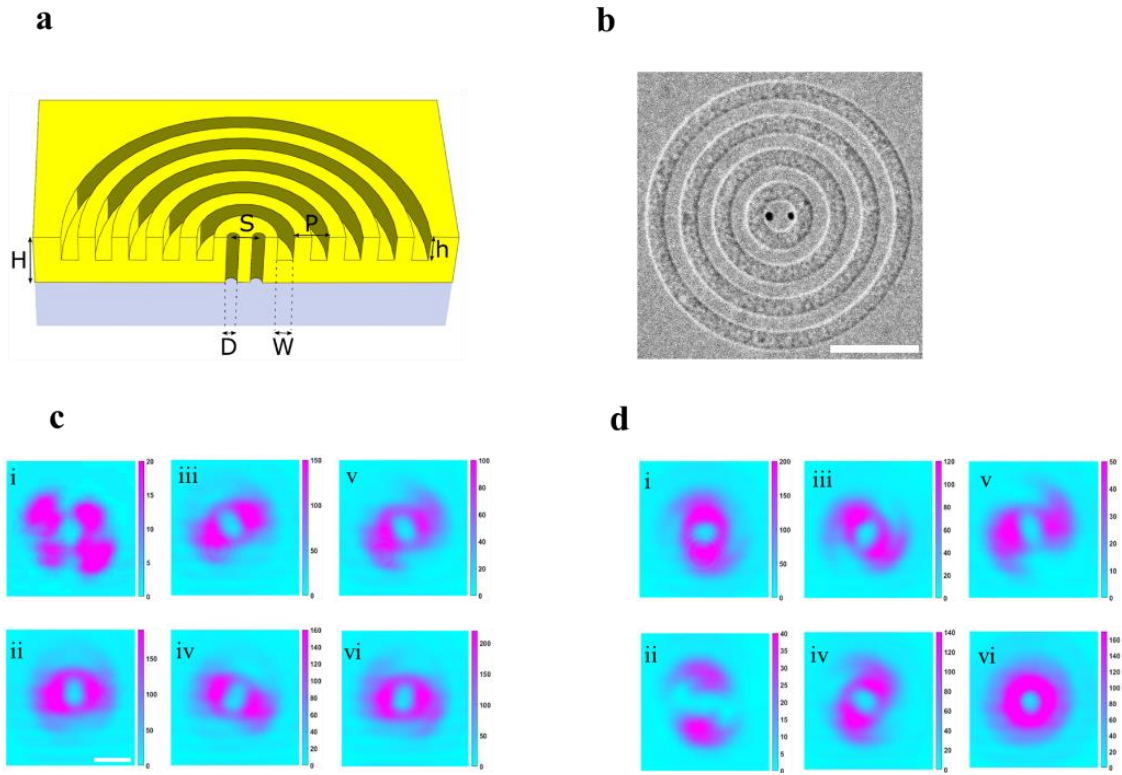

**Figure 1 | Device structure and Stokes Parameters. a)** Schematic of the double holes surrounded by circular grooves. The proposed device is composed of a pair of holes with spacing S (center-to-center spacing between two holes) and diameter D. The plasmonic holes are surrounded by a few periods of circular grating with periodicity P, width W and depth h. The holes and grooves are fabricated on a gold film deposited on glass. SEM image of a typical device. The scale bar in the SEM image is. **b)** SEM image of a device. The scale bar is 2 μm. **c)** and **d)** BFP images corresponding to different configurations of the analyzer under linear and circular polarizations of the incident beam, respectively. The scale bar in **c)** is 280 μm.

The nanoholes have a center-to-center separation of 400 nm and an average aperture diameter of 150 nm. The schematic and scanning electron microscope (SEM) of the device is shown in **Figure 1a** and **Figure 1b**, respectively. The dimensions are verified by SEM images. We use the finite-difference time-domain (FDTD) method through commercial software Lumerical FDTD (Ansys, 2021) to simulate transmission patterns from the bull's eye antenna with double holes, where a plane wave with varying polarization illuminates the structure, and observations are made from the glass side. For circularly polarized light, two orthogonally polarized plane waves with a 90° phase shift are applied. Transmitted fields are collected, and a near-to-far field transformation is performed to analyze the patterns, with Stokes polarimetry used to study the polarization state. BFP microscopy is an imaging technique to investigate the optical properties of light emitted from or transmitted through nanoscale regions. This method, widely used in fields such as biology and photonics, provides detailed information about the transmitted or emitted light, including polarization, intensity, phase, and spatial frequency (momentum), by capturing images at the back focal plane of the objective lens.

To obtain a comprehensive understanding of the polarization state, we perform the Stokes parameter analysis on all pixels corresponding to wave vectors within the numerical aperture (NA) of the objective lens. The optical testing setup, shown in **Figure S1**, is a transmission microscope using an 800 nm laser focused onto the sample with a long working distance objective. Polarization is controlled with a quarter-waveplate (QWP) and linear polarizer. Transmitted light is collected by a second microscope objective, collimated with a convex lens, and recorded by a CCD to capture the radiation pattern. Polarimetry uses a QWP and polarizer between the collimating lens and a flip mirror. Imaging in real and Fourier spaces is achieved with a plano-convex lens. To measure the four Stokes parameters, we conducted six different experiments with various angles of the QWP and linear polarizer relative to the x-axis.

**Table 1** Measured intensities under various analyzer configurations (QWP angle $\Theta$ and linear polarizer angle $\Phi$).

| Intensity | $\Theta$ (°) | $\Phi$ (°) |
|---|---|---|
| $I_a$ | 0 | 0 |
| $I_b$ | 90 | 90 |
| $I_c$ | 45 | 45 |
| $I_d$ | 135 | 135 |
| $I_e$ | 0 | 45 |
| $I_f$ | 0 | 135 |

With $I_a...I_f$ denoting the transmitted light intensity, the first Stokes parameter, $S_0 = I_a + I_b$, represents the total intensity of the transmitted light. The second parameter, $S_1 = I_a - I_b$, measures the vertical or horizontal polarization of light. The third parameter, $S_2 = I_c - I_d$, indicates whether the linear polarization aligns with ±45° angles. The fourth parameter,

$S_3 = I_e - I_f$, captures the circular polarization component, reflecting the left- or right-handiness of the circular polarization state. **Table 1** outlines the configuration of the polarizer and QWP for the six experiments, where $\Phi$ and $\Theta$ represent the angles of the linear polarizer and QWP relative to the x-axis, respectively.

**Figures 1c and d** present cases under linearly and circularly polarized incident light, respectively. For most experiments, a donut-shaped or double lobe-shaped radiation pattern is observed, which rotated with changes in the angle configuration of the QWP and linear polarizer. Under linearly polarized light illumination, the lowest and highest transmission coefficients are recorded as $I_a$ and $I_f$. For circularly polarized incident light, these values are recorded in $I_e$ and $I_a$, respectively. The collected angular distribution of the beaming of the radiation pattern has been massively studied in the literature and can be explained by the excitation of SPPs, coupling and scattering out by the grating[12,13,16,37].

To understand how our design of plasmonic double holes embedded in a bull's eye antenna affects the polarization states and radiation pattern, we numerically compare the optical response of a bare plasmonic double holes to that of a double hole encompassed by circular grooves, under various incident polarization conditions. To quantify the polarization state of the transmitted light, a near-to-far field transformation is applied to the electromagnetic field data collected by a planar monitor. The Stokes parameters are then calculated using the standard formulations provided in **Equation 1**, which offer a full description of the polarization state of the transmitted light in terms of intensity and phase relationships between orthogonal field components.

$$S_0 = E_x^2 + E_y^2$$
$$S_1 = E_x^2 - E_y^2$$
$$S_2 = E_x E_y^* + E_y E_x^*$$
$$S_3 = iE_x E_y^* - iE_y E_x^*$$

**Equation 1**

**Figure 2a** and **c** depict the simulated optical response of the bare double holes under circularly and linearly polarized incident radiation, respectively. Analysis of the Stokes parameters $S_3$ and $S_1$ reveals that the transmitted light largely preserves the polarization state of the incident light, with other Stokes parameters being close to zero across all angles in the Fourier space, indicating minimal polarization conversion. In contrast, in our design, **Figure 2b** and **d** demonstrate that the antenna induces significant polarization modifications (**Figure S2** provides more details). The $S_3$ parameter identifies regions in the Fourier space where the handedness of the incident beam is completely reversed, forming a ring-shaped pattern of opposite helicity. Furthermore, non-zero values of $S_1$ and $S_2$ in specific regions indicate a complete conversion from circular to linear polarization. **Figure 2d** illustrates the transformation of linearly polarized incident light to circularly polarized transmitted light. These observations provide evidence of strong interaction of light with the bull's eye, resulting in substantial alterations in the polarization state of the transmitted light.

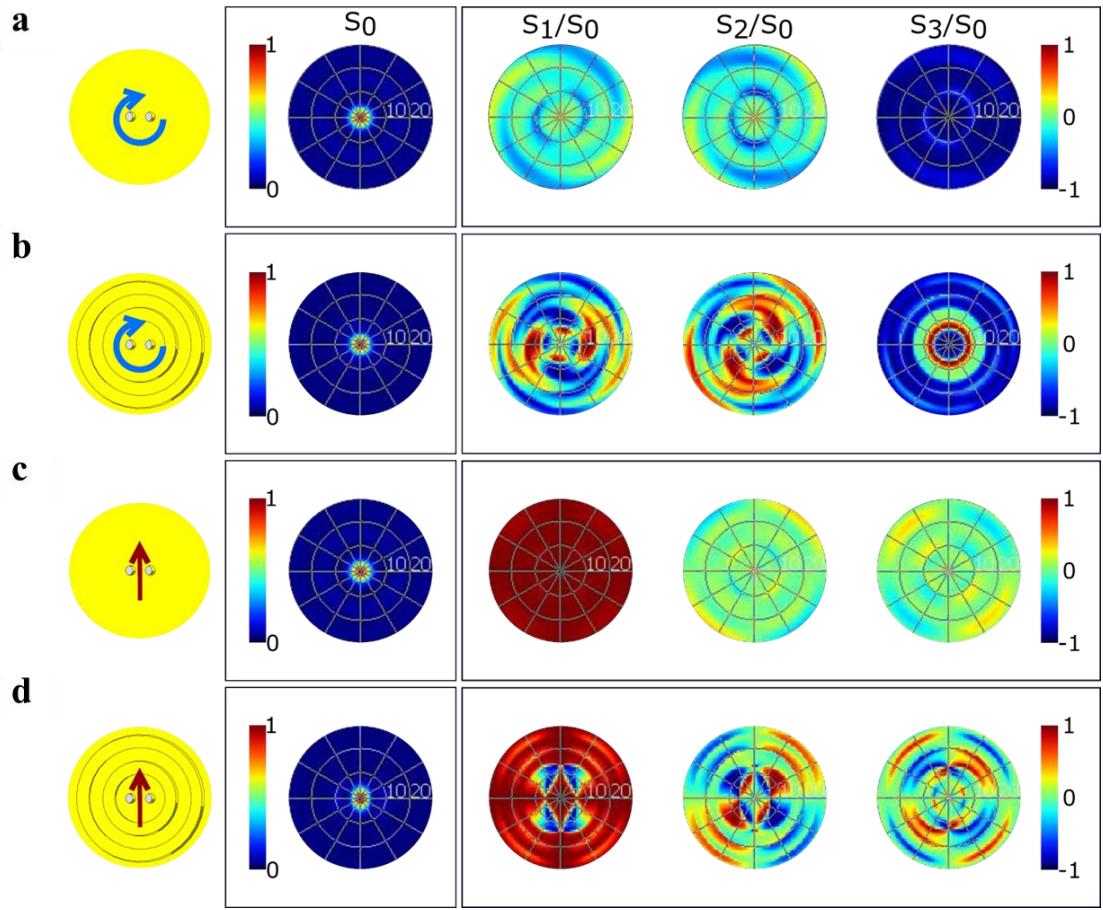

**Figure 2 | Simulation of the bare double holes pair vs the double holes embedded in the bull's eye structure. a) and c)** the normalized Stokes parameters for the bare double holes under circular and linear polarizations of the incident beam, as shown schematically by blue and red arrows. **b) and d)** normalized Stokes parameters of the bull's eye with the double holes under circular and linear polarizations of the incident beam. Here, hole separation and diameter are 200 nm and 100 nm, respectively. Light illuminates the structure from the top, and the thickness of the gold is 200 nm.

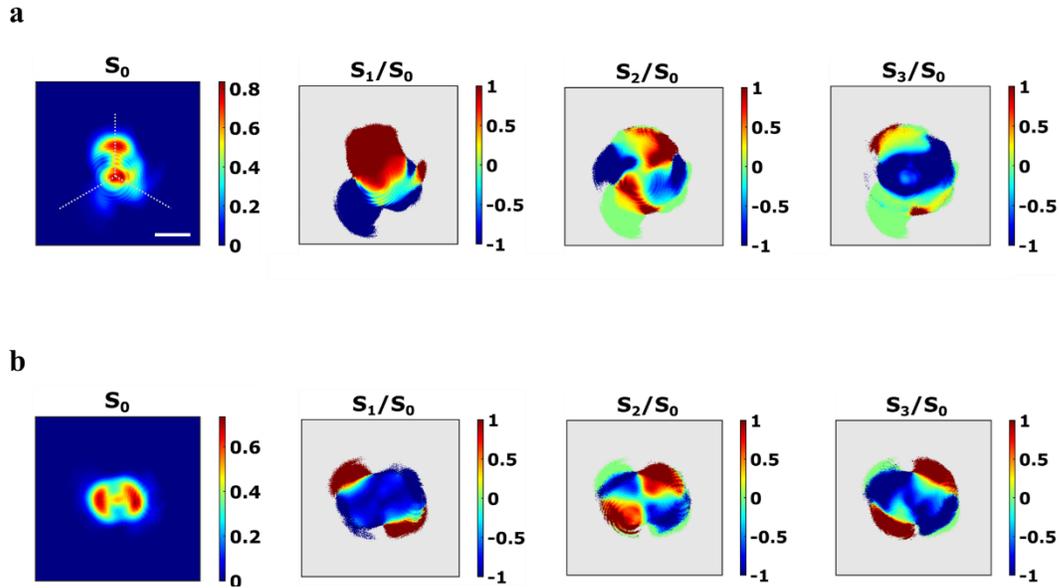

**Figure 3 | Experimental Polarimetric by BFP Imaging of bull's eye with double holes embedded.** Normalized Stokes parameters of any point in the Fourier space within the NA of the microscope objective, **a)** under circular polarization of the incident beam. The dotted lines highlight the subtle three-fold rotational symmetry **b)** under linear polarization of the incident beam. The periodicity and width of the groove are 700 nm and 350 nm, respectively. The scale bar in **c)** is 300 μm.

To experimentally investigate the polarization conversion properties of our structure, we perform polarimetric BFP imaging through Stokes parameter analysis. Our structure comprises a pair of holes with 400 nm spacing, surrounded by six concentric grooves with a periodicity of 700 nm, under circularly polarized incident radiation, **Figure 3a** depicts that the Stokes parameter $S_0$, which represents the total intensity, exhibits a subtle three-fold rotational symmetry in its low-intensity regions, highlighted by the dotted lines in **Figure 3a**, $S_0$. This pattern contrasts with the intensity distribution observed under linearly polarized excitation in **Figure 3b**. Additionally, the central region of $S_0$ displays two high-intensity lobes bisected by an area of diminished intensity. Strong three-fold and higher-order rotational symmetries have been observed in chiral plasmonic structures[38]. This complex intensity distribution suggests a non-trivial interaction between the circularly polarized incident field and the plasmonic nanostructure, potentially arising from the interplay between different SPP modes and their coupling to the far-field radiation. In the case of circularly polarized illumination, the $S_1$ parameter, which corresponds to vertical and horizontal state of polarization, reveals pronounced linear polarization components at wave vectors corresponding to the three-fold rotationally symmetric regions. However, the Stokes parameter $S_2$, which represents the degree of polarization aligning with ±45° angles, exhibits a spatial distribution in the Fourier space that closely resembles that observed under linear polarization. Notably, the $S_3$ parameter, which describes the circular state of the polarization, under circular illumination indicates that the central region weakly preserves the incident polarization state. In summary, in both experimental configurations, we observed significant polarization state conversion, particularly pronounced in the central region of the Fourier space.

On the other hand, under linearly polarized illumination, **Figure 3b** shows that the Stokes parameter $S_0$ exhibits an angular distribution characterized by two high-intensity lobes separated by a region of reduced intensity. The distribution of $S_1$ indicates that most of the Fourier space retains the polarization state of the incident light. However, distinct regions display orthogonal polarizations relative to the incident beam, confirming polarization conversion to the orthogonal state. The distribution of $S_2$ reveals a cross-shaped pattern with alternating positive and negative values, suggesting the presence of diagonal linear polarization components. The $S_3$ parameter displays four mutually orthogonal lobes with distinct circular polarization states, including a notable region with circular polarization at normal incidence ($k_x = k_y = 0$). These observations suggest complex polarization conversion processes mediated by the plasmonic nanostructure, potentially involving asymmetric scattering and interference effects that depend on the incident light's polarization state and the geometric parameters of the bull's eye structure.

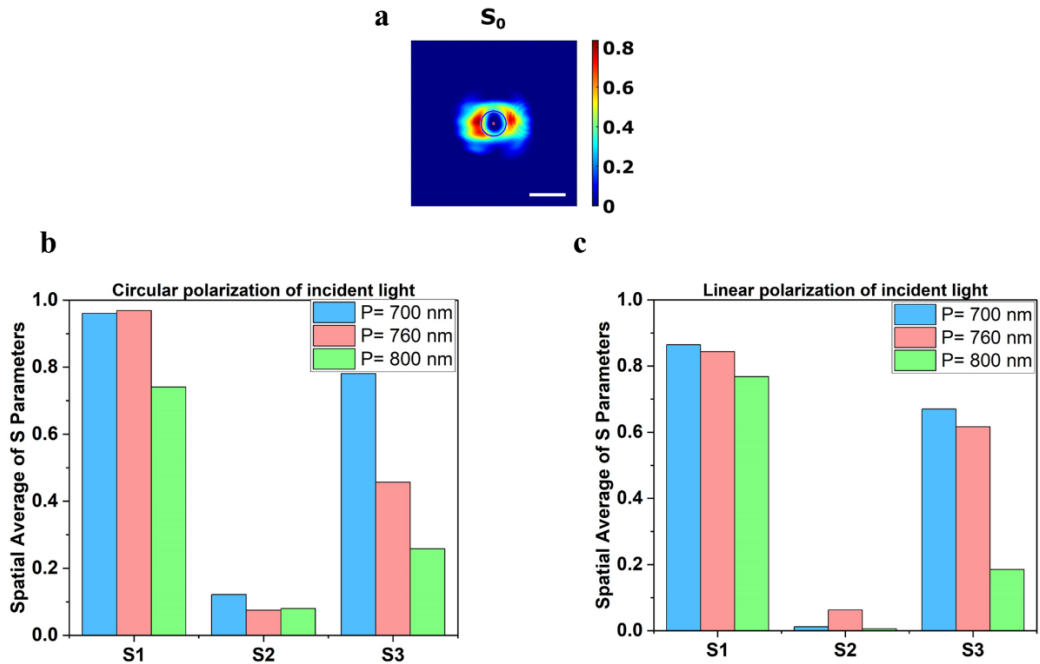

**Figure 4 | Effect of the bull's eye periodicity on the device's performance in polarization conversion. a)** The blue circle in the image indicates the region utilized for averaging the Stokes parameters **b)** The polarization conversion of three devices with periodicities ranging s from700 nm to 800 nm, under circular polarization of the incident beam. **c)** Same as **b)** but under linear vertical polarization of the incident beam.

To quantitatively assess the polarization conversion efficiency and study the impact of the bull's eye antenna periodicity, we focus our analysis on the central region, highlighted by the blue circle in **Figure 4a**, of the Fourier plane ($k_x = k_y = 0$). This region is of particular interest due to its fundamental significance in both theoretical photonics and practical applications like slow light and beaming[12,13,39]. **Figure 4b** presents the area-averaged Stokes parameters when a circularly polarized beam is normally incident on the structure. The comparison of the $S_1$ under different grating periodicities shows high values (close to 1) for all periodicities. This suggests that a significant portion of the transmitted light is converted from circular to linear polarization. On the other hand, small values of $S_2$ for all periodicities imply a negligible conversion to the diagonally linear polarization states. The same quantity is presented in **Figure 4c** when a linearly polarized beam is normally incident on the structure. The values of $S_1$ show that the transmitted light preserves the polarization of incident illumination. However, the values of $S_3$ reveal noticeable linear to circular polarization conversion for all periodicities.

**Conclusion**

The study highlights the sophisticated polarization conversion properties of a plasmonic double-hole surrounded by a bull's eye antenna, examined through a combination of FDTD simulation and experimental BFP microscopy along with Stokes parameter analysis. This work reveals that the structure is capable of significantly altering the polarization state of the incident light, namely, the conversion from linear to circular polarization and vice versa. These transformations are highly dependent on the periodicity of the bull's eye geometry, with notable effects in the central region of the Fourier space.

The research observed asymmetric scattering and distinct Stokes parameter patterns across various incident polarization states. These results emphasize the bull's eye with plasmonic double holes' ability to manipulate light polarization at the nanoscale, which opens new avenues for developing polarization-controlled emitters and sensors. Moreover, this work contributes to a deeper understanding of light-matter interaction in plasmonic nanostructures, which may pave the way to potential applications in quantum optics and biomedical sensing. Future research could focus on optimizing the geometry for specific polarization tasks and exploring its wavelength dependence.

AUTHOR INFORMATION

**Corresponding Author**

**Qing Gu** - Department of Electrical and Computer Engineering, North Carolina State University, Raleigh, 27695, USA and Department of Physics, North Carolina State University, Raleigh, 27695, USA.; Email: qgu3@ncsu.edu

**Authors**

**Abbas Ghaffari**- Department of Electrical and Computer Engineering, North Carolina State University, Raleigh, 27695, USA.

**Somayeh Kashani** - Department of Electrical and Computer Engineering, North Carolina State University, Raleigh, 27695, USA.

**Jiazhen Li** - Department of Electrical and Computer Engineering, North Carolina State University, Raleigh, 27695, USA.

**Paschalis Gkoupidenis**- - Department of Electrical and Computer Engineering, North Carolina State University, Raleigh, 27695, USA and Department of Physics, North Carolina State University, Raleigh, 27695, USA.

**Robert Riehn** - Department of Physics, North Carolina State University, Raleigh, 27695, USA.

**Author Contributions**

A.G. conceptualized the research, fabricated the devices, and conducted both FDTD simulations and experimental work. S.K. contributed to data analysis, image processing, and manuscript preparation. J.L. provided technical assistance with the optical setup. P.G. offered additional support. R.R. and Q.G. supervised the project, guided the development of ideas, and assisted in manuscript drafting.

**Notes**

The authors declare no conflict of interest.

ACKNOWLEDGMENT

Q. Gu acknowledges support from the National Science Foundation (CAREER ECCS-2209871, ECCS-2240448). R. Riehn acknowledges support from the National Institutes of Health (GM126887) and the US Air Force Office of Scientific Research (AFOSR, FA9550-23-1-0311).